\documentclass[10pt,journal, twocolumn]{IEEEtran}
\usepackage{graphicx,times, amsmath, amsfonts,comment}
\usepackage{amssymb,epstopdf,bm,bbm,amsthm}
\usepackage[noend]{algorithmic}

\usepackage{multirow, mathtools}

\usepackage{epstopdf, soul, color}
\usepackage{algorithm, array}

\usepackage{relsize}
\newcommand{\beq}{\begin{equation}}
\newcommand{\eeq}{\end{equation}}

\theoremstyle{plain}

\begin{document}

\title{Coordinated Multi-Point (CoMP) Transmission in Downlink Multi-cell NOMA Systems: Models and Spectral Efficiency Performance}
\author{\IEEEauthorblockN{Md Shipon Ali, Ekram Hossain, and Dong In Kim}
\thanks{M. S. Ali and E. Hossain are with the Department of Electrical and Computer Engineering, at the University of Manitoba, Canada (Emails: alims@myumanitoba.ca, ekram.hossain@umanitoba.ca). D. I. Kim is with the School of Information and Communication
Engineering at the Sungkyunkwan University (SKKU), Korea (email: dikim@skku.ac.kr). 
The work was supported by a Discovery Grant form the Natural Sciences and Engineering Research Council of Canada (NSERC) and in part by  the National Research Foundation of Korea (NRF) grant funded by the Korean government (MSIP) 
(2017R1A2B2003953 and 2014R1A5A1011478).
} 
}
\maketitle

\begin{abstract}
We outline a general framework to use coordinated multi-point (CoMP) transmission technology in downlink multi-cell non-orthogonal multiple access (NOMA) systems considering distributed power allocation at each cell. In this framework, CoMP transmission is used for users experiencing strong receive-signals from multiple cells while each cell adopts NOMA for resource allocation to its active users. After a brief review of the working principles of different CoMP schemes, we identify their applicability and necessary conditions for their use in downlink multi-cell NOMA system. After that, we discuss different network scenarios with different spatial distributions of users and study the applicability of CoMP schemes in these network scenarios. To the end, a numerical performance evaluation is carried out for the proposed CoMP-NOMA systems and the results are compared with those for conventional orthogonal multiple access (OMA)-based CoMP systems. The numerical results quantify the spectral efficiency gain of the proposed CoMP-NOMA models over CoMP-OMA. Finally, we conclude this article by identifying the potential major challenges in implementing CoMP-NOMA in future cellular systems.
\end{abstract}


\section*{Introduction}
Recently, non-orthogonal multiple access (NOMA\footnote{In this article, NOMA refers to the power domain NOMA.}) has received tremendous interests from both academia and industry. Due to its potential to significantly enhance the spectral efficiency of transmission, NOMA is being considered as a promising multiple access technology for fifth generation (5G) and beyond 5G (B5G) cellular systems \cite{saito2013}-\cite{ding2017}. The fundamental idea of NOMA  is to simultaneously serve multiple users over the same spectrum resources (time, frequency, and space) at the expense of inter-user interference. In downlink NOMA, a base station (BS) transmitter schedules multiple users in the same spectrum resources and superposes all users' signals into power domain by exploiting their respective channel gains, while successive interference cancellation (SIC) is applied at the users' receiver ends for inter-user interference cancellation.

To maximize the overall spectral efficiency and/or minimize the total power consumption in downlink NOMA, the BS allocates transmit power in such a way that the SIC decoding is performed according to an ascending order of the channel gains of the NOMA users \cite{msali2016}. 
This power allocation strategy results in a low  received signal-to-intra-cell-interference ratio for lower channel gain users (e.g., cell-edge users) who are also vulnerable to inter-cell interference. Therefore, inter-cell interference management will be crucial in multi-cell downlink NOMA systems. To mitigate inter-cell interference for traditional downlink orthogonal multiple access (OMA)-based 4G cellular systems, third generation partnership project (3GPP) adopted CoMP transmission technique in which multiple cells, called \textit{CoMP set}, coordinately schedule/transmit to the interference-prone users \cite{3GPP2011}-\cite{online2014}. 
In this article, we focus on the application of CoMP in NOMA-based multi-cell downlink transmission scenarios in order to improve the spectral efficiency performance of the system.

Recently, some studies have investigated on combining CoMP with NOMA in downlink transmission scenarios. The author in \cite{choi2014} utilized Alamouti code for joint transmission to a cell-edge user in a two-cell CoMP set. In this work, $2$-user NOMA is used in both the cells where the non-CoMP user is the cell-center user. Another work on downlink CoMP-NOMA can be found in \cite{tian2016}, where an opportunistic CoMP-NOMA system is used for a group of cell-edge users which receive strong signals from all the coordinating cells. In this work, a joint multi-cell power allocation strategy is used based on the CoMP users'  channel gains. In \cite{bey2016}, a downlink CoMP-NOMA system was studied considering multiple antennas at transmitter and receiver ends. Moreover, a downlink multi-cell NOMA application can be found in \cite{shin2016} where the authors considered only one CoMP-user\footnote{In this article, a user requiring CoMP transmission is referred to as a CoMP-user, while a user who does not require a  CoMP transmission is referred to as a non-CoMP-user.} group with one non-CoMP-user at each NOMA cluster among the coordinating cells.
 
Different from the aforementioned works, we utilize distributed power allocation for NOMA users in each cell, while various CoMP schemes are applied to the cell-edge users experiencing inter-cell interference. In our model, we first determine the users requiring CoMP transmissions from multiple cells and those requiring single transmissions from their serving cells. After that, different NOMA clusters are formed in individual cells in which the CoMP-users are clustered with the non-CoMP-users in a NOMA cluster. After clustering the users, each cell independently allocates transmit power to its NOMA users by utilizing a dynamic power allocation method \cite{msali2016} for cell sum-throughput maximization. However, in a CoMP-NOMA system, the decoding order for the users in a NOMA cluster will be different from that in a conventional single-cell NOMA system as in \cite{msali2016}, and thus the power allocation solution  will also be different although it can be solved in a way similar to that in \cite{msali2016}. In addition, we consider single antenna at transmitter and receiver ends. However, in case of MIMO, this model is also valid while MIMO-NOMA beamforming and power allocation \cite{msali2017}-\cite{higuchi2015} needs to consider the CoMP effects. 

The rest of the article is organized as follows. We first discuss various CoMP schemes and identify their applicability and necessary conditions to apply them in downlink multi-cell NOMA systems. After that, we present various CoMP-NOMA deployment approaches and the limitations and potential gains for each of the approaches. Numerical results based on simulations of the proposed approaches are presented and compared with those for the CoMP-OMA approaches. Finally, we conclude this article by identifying potential challenges to implement downlink CoMP-NOMA systems.

\section*{CoMP Schemes for Downlink CoMP-NOMA}
We will first define the achievable throughput for a NOMA user according to their decoding order in a downlink NOMA system. Then we will discuss different CoMP schemes considering single antenna BS and user equipment (UE), and identify their applicability for a NOMA-based transmission model. 


\subsection*{Achievable Downlink Throughput for a NOMA User}

Let us assume a downlink NOMA cluster with $n$ users and the following decoding order: UE$_1$ is decoded first, UE$_2$ is decoded second, and so on. Therefore, UE$_1$'s signal will be decoded at all the users' ends, while UE$_n$'s signal will be decoded only at her own end. Since UE$_1$ can only decode her own signal, it experiences all the other users' signals as interference, while UE$_n$ can decode all users' signals and removes inter-user interference by applying SIC. Therefore, the achievable throughput for the $i$-th user can be written as follows: $R_i = B \log_2\Bigg(1+\frac{p_i \gamma_i}{\sum\limits_{j = i+1}^{n} p_j\gamma_i+1}\Bigg), \, \, \forall i = 1,2,\cdots,n$, where $\gamma$ is the normalized channel gain with respect to noise power density over NOMA bandwidth $B$, and $p_i$ is the allocated transmit power for UE$_i$. The necessary condition for power allocation to perform SIC is $\Big(p_i - \sum\limits_{j=i+1}^{n} p_j\Big)\gamma_j\geq p_{tol}, \, \, \forall i = 1,2,\cdots,n-1$, where $p_{tol}$ is the minimum difference in received power (normalized with respect to noise power) between the decoded signal and the non-decoded inter-user interference signals \cite{msali2016}. To maximize the overall cell throughput and/or minimize the power consumption, the decoding should be performed in an ascending order of the channel power gains of the NOMA users. That is, the aforementioned decoding order will provide maximum sum-throughput if the channel gain for NOMA users are such that: $\gamma_n > \gamma_{n-1}>\cdots> \gamma_1$. In such a optimal scenario, the power allocation condition could be simplified as $\Big(p_i - \sum\limits_{j=i+1}^{n} p_j\Big)\gamma_i\geq p_{tol}, \, \, \forall i = 1,2,\cdots,n-1$ \cite{msali2016}.

\begin{figure}[h]
\begin{center}
	\includegraphics[width=3.4 in]{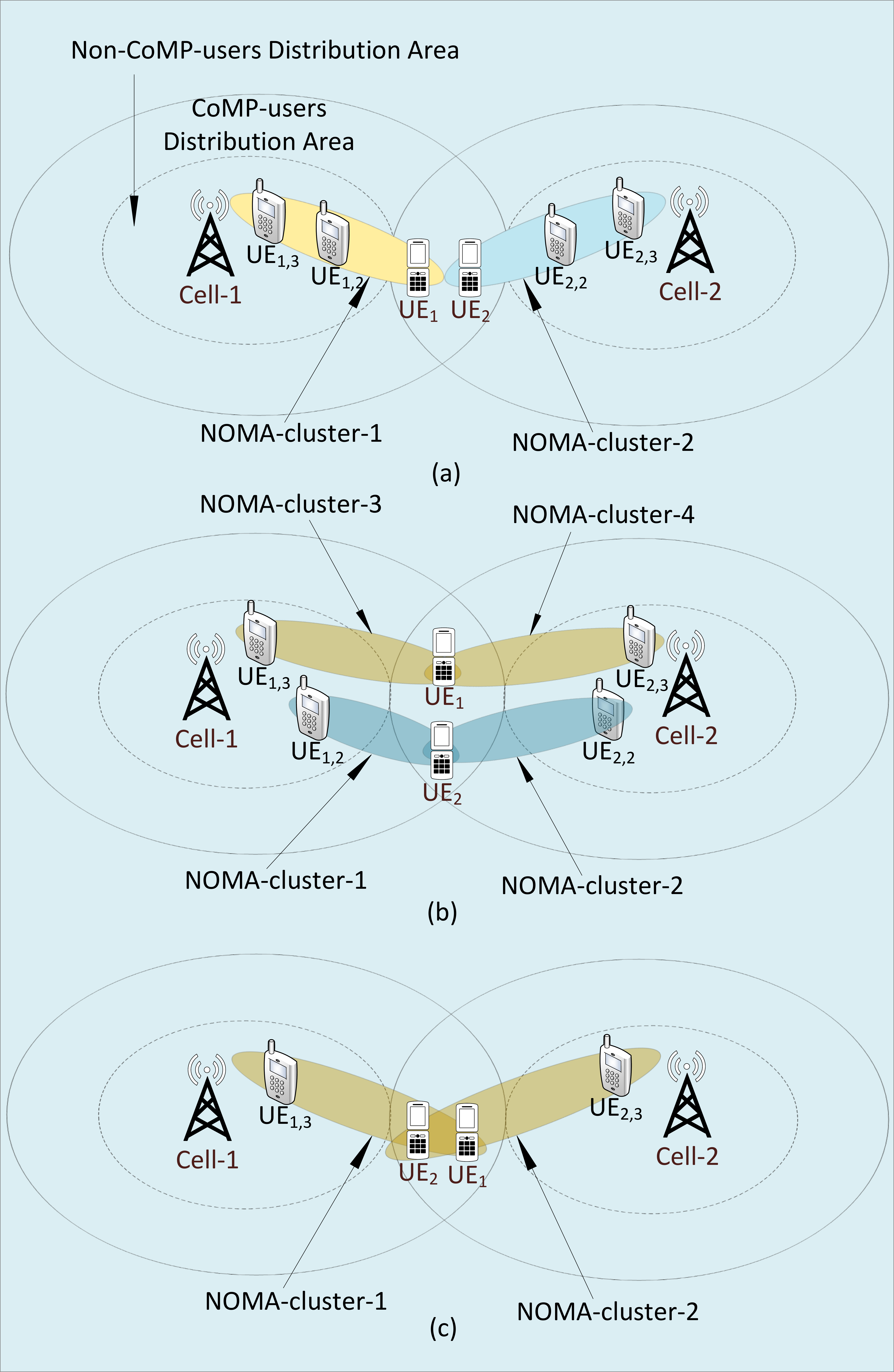}
	\caption{Illustrations of the various CoMP schemes for a downlink NOMA system: (a) CS-CoMP-NOMA, (b) JT-CoMP-NOMA for multiple CoMP-users and multiple non-CoMP-users, and (c) JT-CoMP-NOMA for multiple CoMP-users and a single non-CoMP-user.}
	\label{fig:i}
 \end{center}
\end{figure} 
\begin{figure}[h]
\begin{center}
	\includegraphics[width=3.4 in]{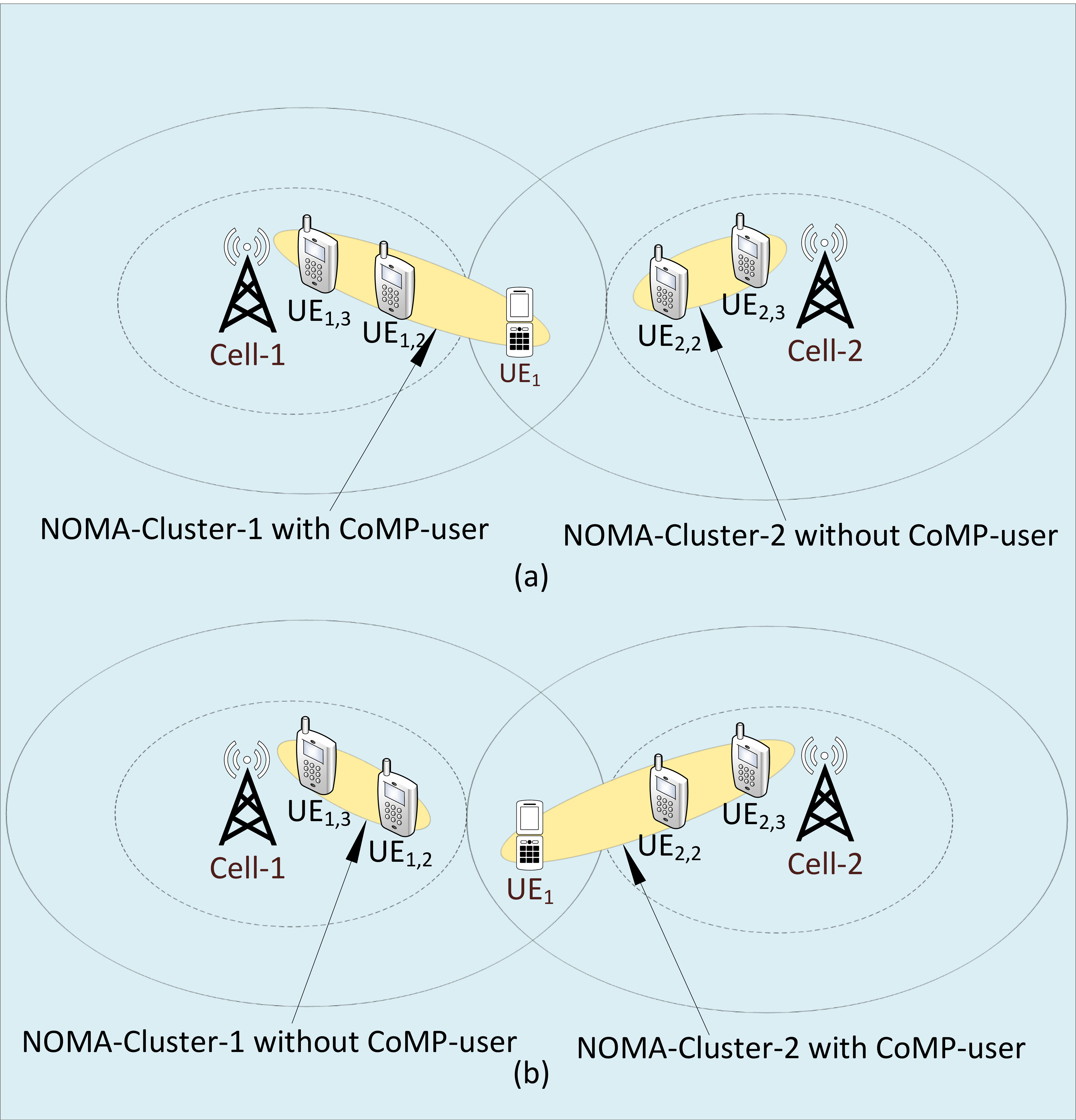}
    \caption{Illustrations of the DPS-CoMP schemes at downlink NOMA system for a single CoMP-user: (a) CoMP-user is clustered in cell$-1$, (b) CoMP-user is clustered in cell$-2$.}
	\label{fig:ii}
 \end{center}
\end{figure} 

\subsection*{Coordinated Scheduling (CS)-CoMP in Downlink NOMA Systems} 
In CS-CoMP, CoMP-users are scheduled on orthogonal spectrum resources and receive desired signals only from their serving cells, respectively, while an orthogonal spectrum allocation is done based on coordination among the CoMP-cells. In CS-CoMP-NOMA, each CoMP-user is grouped into one NOMA cluster and does not experience inter-cell interference due to the orthogonal spectrum allocation among the CoMP-cells. Therefore, the working principle of CS-CoMP-NOMA is similar to that of traditional single-cell NOMA. 

Fig. \ref{fig:i}(a) illustrates CS-CoMP-NOMA for a homogeneous two-cell CoMP scenario, where two CoMP-users, UE$_1$ and UE$_2$, are allocated orthogonal spectrum resources from cell$-1$ and cell$-2$, respectively. In cell$-1$, UE$_1$ is grouped into a NOMA cluster with the non-CoMP-users UE$_{1,2}$ and UE$_{1,3}$ of cell$-1$. Similarly, UE$_2$ forms a NOMA cluster with the non-CoMP-users UE$_{2,2}$ and UE$_{2,3}$ at cell$-2$. After spectrum resources are allocated to each NOMA cluster, the CS-CoMP-NOMA is similar to the traditional single-cell NOMA which offers spectral efficiency gains over its OMA counterparts \cite{msali2016}.

\subsection*{Joint Transmission (JT)-CoMP in Downlink NOMA Systems}
In JT-CoMP schemes under single antenna BS and UE, multiple cells (i.e., CoMP-cells) simultaneously transmit the same data to a CoMP-user by using the same spectrum resources \cite{online2014}. Since the same data is sent by all CoMP-cells, the reception performance for CoMP-users is improved. In a JT-CoMP-OMA system with multiple CoMP-users served by the same CoMP set, the users are scheduled on orthogonal spectrum resources. However, in JT-CoMP-NOMA, one or multiple non-CoMP-users from each cell form NOMA cluster with one or multiple CoMP-users. 

In Fig. \ref{fig:i}(b), a JT-CoMP-NOMA scheme is illustrated where two CoMP-users are grouped into two different NOMA clusters with different non-CoMP users. Thus, UE$_1$ receives the same message from cell$-1$ and cell$-2$ simultaneously over the same spectrum resource which is orthogonal to that for the other CoMP-user (i.e., UE$_2$).  On the other hand, in Fig. \ref{fig:i}(c), two CoMP-users are in the same NOMA cluster and both of them receive transmissions from the two CoMP cells simultaneously over the same spectrum resource. The CoMP-users in a NOMA cluster receive their desired signals by utilizing SIC  according to their decoding order. However, in a JT-CoMP-NOMA system, for successful decoding in presence of multiple CoMP-users in a NOMA cluster, the two following necessary conditions need to be satisfied: 

\begin{itemize}
\item \textbf{The signals for users receiving CoMP transmissions will be decoded prior to those for the users receiving single transmissions from their serving cells}. To illustrate this condition, let us consider Fig. \ref{fig:i}(c), where two CoMP-users, UE$_1$ and UE$_2$, receive CoMP transmission from both cells, while two non-CoMP-users, UE$_{1,3}$ and UE$_{2,3}$, receive their desired signals only from cell$-1$ and cell$-2$, respectively. Now, with the JT-CoMP scheme, the message signals for UE$_1$ and UE$_2$ will need to be decoded prior to decoding the  signals for UE$_{1,3}$ and UE$_{2,3}$. In other words, UE$_{1,3}$ and UE$_{2,3}$ will  decode and cancel the message signals for UE$_1$ and UE$_2$ prior to decoding their own signals. To verify the condition, let us consider the opposite scenario, that is, UE$_1$ and UE$_2$ need to decode the message signals for UE$_{1,3}$ and UE$_{2,3}$ before decoding their desired signals. To decode UE$_{1,3}$'s signal at UE$_1$ or UE$_2$ end, the received power for UE$_{1,3}$ need to be higher than the summation of the received powers for UE$_1$ and UE$_2$. Note that,  UE$_1$,  UE$_2$, and UE$_{1,3}$ are in the same NOMA cluster. Although cell$-1$ can allocate more power for UE$_{1,3}$ than the sum power for UE$_1$ and UE$_2$, the received power for UE$_{1,3}$ cannot  be guaranteed to be higher than the sum of the received powers for UE$_1$ and UE$_2$. The reason is that both UE$_1$ and UE$_2$ will receive the same signal from both of the CoMP-cells, and thus their received powers will be improved. 


\item \textbf{The decoding order for a CoMP-user will be same in all NOMA clusters formed at different CoMP-cells in which the CoMP-user is clustered}. To illustrate this condition, let us again consider Fig. \ref{fig:i}(c). If the decoding order for UE$_1$ is prior than UE$_2$ in cell$-1$, then it would be similar  for cell$-2$ regardless the channel gains of UE$_1$ and UE$_2$ in cell$-2$. SIC is only possible at CoMP-user ends if this condition is satisfied. This condition also implies that the traditional power allocation for cell-throughput maximization will not hold  in a JT-CoMP-NOMA system.
\end{itemize}

\subsection*{Dynamic Point Selection (DPS)-CoMP in Downlink NOMA Systems}
In a DPS-CoMP system, the data streams for each CoMP-user  become available in all of the CoMP-cells but only one cell sends data at a time. In each subframe, all the CoMP-cells check the channel quality for each CoMP-user, and based on the maximum channel gain only one cell is dynamically selected for data transmission. Therefore, DPS-CoMP can be applied in a NOMA system where the user clustering and power allocation need to be performed at each subframe. After determining the serving cell in DPS-CoMP, a CoMP-user is grouped into a NOMA cluster with the non-CoMP-users served by that cell. 

Fig. \ref{fig:ii}(a) and fig. \ref{fig:ii}(b) illustrate the working principle of DPS-CoMP-NOMA at two different subframes, by assuming that the CoMP-user has better channel gain with cell$-1$ in subframe$-1$ and with cell$-2$ in subframe$-2$.  Since each CoMP-user is grouped into one cluster at a time and does not experience any inter-cell interference, the decoding order and power allocation for DPS-CoMP-NOMA is exactly similar to that for the convention NOMA for a single cell system with dynamic power allocation \cite{msali2016}. 



\subsection*{Coordinated Beamforming (CB)-CoMP in Downlink NOMA Systems}

The fundamental principle of coordinated beamforming (CB)-CoMP is similar to that of the distributed MIMO system, where the coordinating cells act as a distributed antenna array under a virtual BS. One CoMP-user is associated with one CoMP-cell, while all the CoMP-cells use same spectrum resources to serve their associated CoMP-users by utilizing the distributed MIMO principle \cite{online2014}. To apply CB-CoMP scheme in downlink NOMA system, one or multiple non-CoMP-users need to be clustered with a CoMP-user at each CoMP-cell. However, to cancel the inter-cell interference for CoMP-users  using the same spectrum resources, the zero-forcing MIMO beamforming need to be performed by using the CoMP-users' channel vector corresponding to the CoMP-cells. Since the same beam will be used for all non-CoMP-users and a CoMP-user in a CB-CoMP-NOMA cluster, the non-CoMP-user may not be able to decode the message signals due to the mismatch in dimension  between their channel vector (which has single dimension since only one channel exists with the serving cell) and precoding vector (which has a dimension equal to the CoMP-set size, and precoding is done based on the CoMP-users' channel gains). Therefore, CB-CoMP is not applicable in for a CoMP-NOMA system.

\section*{Deployment Scenarios for CoMP in Downlink NOMA Systems}
In this section, we discuss different CoMP-NOMA deployment models based on the users' distribution (i.e., number of CoMP-users and non-CoMP-users) in a two-cell CoMP set.

\begin{figure}[h]
\begin{center}
	\includegraphics[width=3.4 in]{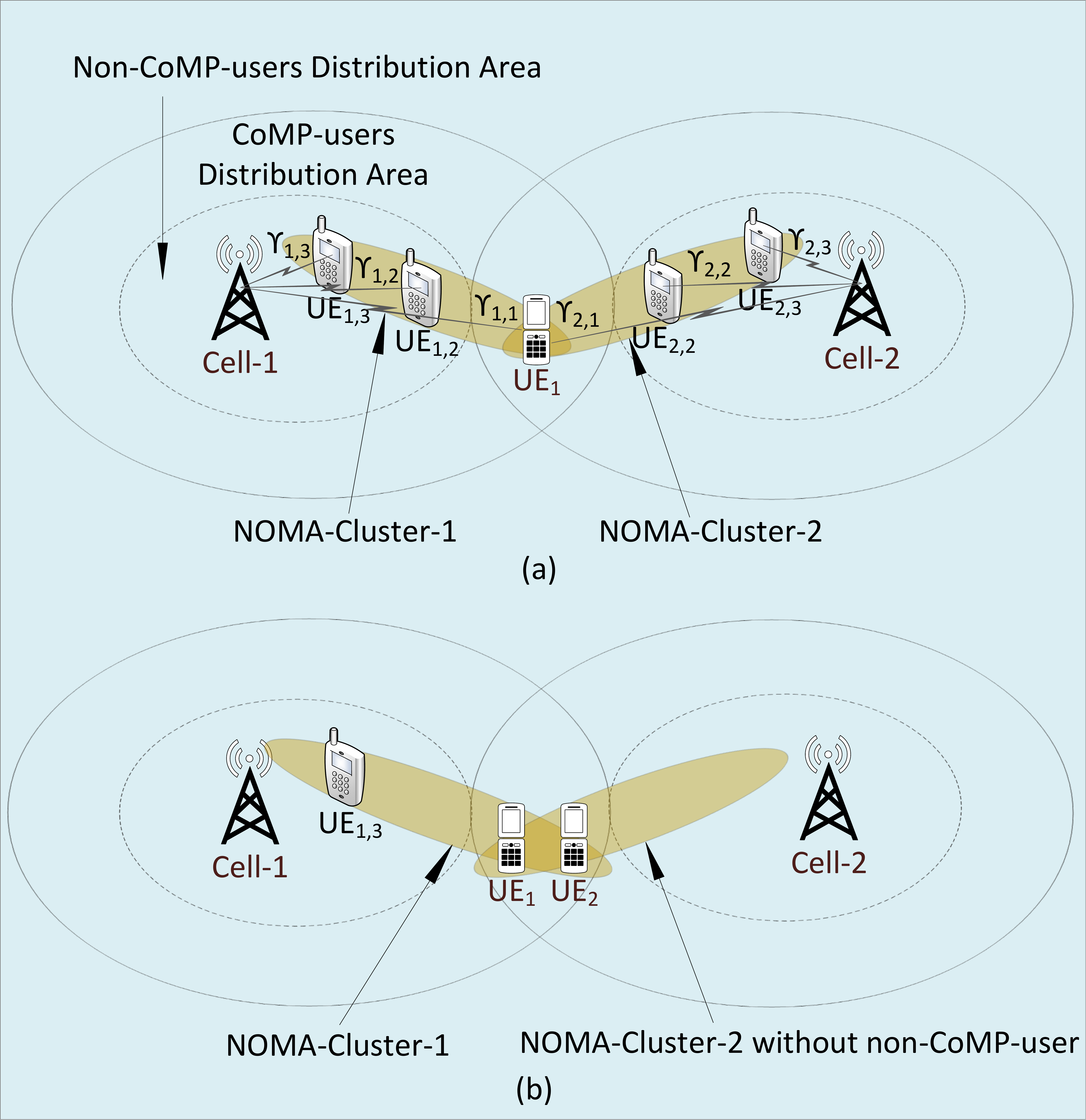}
	\caption{Illustrations of the various CoMP-NOMA deployment scenarios: (a) deployment scenario$-1$, (b) deployment scenario$-3$.}
	\label{fig:iii}
 \end{center}
\end{figure}

\subsection*{CoMP-NOMA Deployment Scenario$-$1}
In this  scenario, only one CoMP-user is considered for a CoMP-set, while one or multiple non-CoMP-users are considered in each CoMP-cell of that CoMP-set. Since there is only one CoMP-user, the JT-CoMP-NOMA and DPS-CoMP-NOMA schemes are applicable in this scenario. Fig. \ref{fig:iii}(a) illustrates the deployment scenario$-1$ for a two-cell CoMP set in a JT-CoMP-NOMA set up, where the CoMP-user is included in both of the NOMA clusters formed in cell$-1$ and cell$-2$, and utilizes the same spectrum resources. By exploiting the NOMA principle, each cell superposes their NOMA users' message signals in the same spectrum resources, and thus the CoMP-user's message signal is superposed at both cells. To decode the desired signal, the decoding order for the CoMP-user needs to be same in both the NOMA clusters. 

Let $\gamma_{1,1}$, $\gamma_{1,2}$, and $\gamma_{1,3}$ denote the normalized channel power gains (with respect to noise power) for UE$_{1}$, UE$_{1,2}$ and UE$_{1,3}$ in cell$-1$, respectively, and $\gamma_{2,1}$, $\gamma_{2,2}$, and $\gamma_{2,3}$ are the normalized channel power gains  for UE$_{1}$, UE$_{2,2}$ and UE$_{2,3}$ in cell$-2$, respectively. If the decoding order is based on the user's subscript, i.e., the message signal for UE$_1$ is decoded prior to decoding the other users' signals in NOMA cluster$-1$ and NOMA cluster$-2$, then the achievable throughput for the CoMP-user is $ R_1 = B \log_2\Bigg(1+ \frac{ \sum\limits_{i=1}^{2}p_{i,1}\gamma_{i,1}}{\sum\limits_{i=1}^{2}\sum\limits_{j=2}^{3} p_{i,j}\gamma_{i,1} + 1} \Bigg)$. The achievable throughput for the $j$-th non-CoMP-user in cell$-i$ is, $R_{i,j} = B \log_2\Big(1+\frac{p_{i,j} \gamma_{i,j}}{\sum\limits_{k=j+1}^{3} p_{i,k}\gamma_{i,j}  + \sum\limits_{m = 1, m\neq i}^{2}\sum\limits_{l = 2}^{3} p_{m,l}\gamma_{i,j}^\prime + 1}\Big)$, where $i = 1,2$ and $j = 2,3$. The term $\gamma_{m,j}^\prime$ is the normalized channel power gain  for the $j$-th non-CoMP-user in $i$-th cell but measured from $m$-th cell ($m\neq i$) of the CoMP set, and represents the inter-cell interference for non-CoMP-users. If the inter-cell interference for a non-CoMP-user from any cell of a CoMP set is negligible, then the achievable NOMA throughput for the non-CoMP users can be approximated as $R_{i,j} = B \log_2\Big(1+\frac{p_{i,j} \gamma_{i,j}}{\sum\limits_{k=j+1}^{3} p_{i,k}\gamma_{i,j} + 1}\Big), \,\, \forall i = 1,2, \text{ and } \forall j = 2,3$. 

\subsection*{CoMP-NOMA Deployment Scenario$-$2}
In this scenario, we assume multiple CoMP-users in a CoMP set, while one or multiple non-CoMP-users in each of the CoMP-cells of that CoMP set. This scenario is similar to those models illustrated in Fig. 1, in which all of the CS-CoMP, JT-CoMP, and DPS-CoMP schemes are applicable. In case of JT-CoMP-NOMA for Fig. \ref{fig:i}(b), the achievable throughput formulas for CoMP-user and non-CoMP users for each NOMA cluster pair, which includes a common CoMP-user, are similar to those for scenario$-1$. For Fig. \ref{fig:i}(b), it is noted that each NOMA cluster can only include one CoMP-user, thus the spectrum resource for different CoMP-users are orthogonal. However, for the JT-CoMP-NOMA deployment scenario$-2$ in Fig. \ref{fig:i}(c), multiple CoMP-users are grouped into each NOMA cluster formed at different CoMP-cells but their decoding order will be similar in all cases. Similar to the scenario$-1$, if the decoding order is based on the user's subscript, then the achievable throughput formula for CoMP-users can be expressed as $ R_j = B \log_2\Bigg(1+ \frac{ \sum\limits_{i=1}^{2}p_{i,j}\gamma_{i,j}}{\sum\limits_{i=1}^{2}\sum\limits_{k=j+1}^{3} p_{i,k}\gamma_{i,j} + 1} \Bigg) \,\, \forall j = 1,2$. On the other hand, the formulas for achievable throughput for non-CoMP users are similar to those for scenario$-1$.

\subsection*{CoMP-NOMA Deployment Scenario$-$3}

In a user-centric CoMP system, different CoMP-users of a particular cell can receive CoMP transmissions from CoMP-cells belonging to different CoMP sets. In such a case, for a JT-CoMP-NOMA system, the CoMP-users of different CoMP sets will interfere with each other  and thus will not form NOMA cluster. Although they can form NOMA cluster by maintaining their decoding order requirement, the inter-cell interference that we have neglected in scenario$-1$ would be excessively high. Therefore, it can be recommended that NOMA clusters are formed by including CoMP-users from one CoMP set at a time. 

Based on the aforementioned idea for CoMP-NOMA clustering, for a particular CoMP-set, some cells may not have non-CoMP-user to form NOMA cluster with the CoMP-users, while other cells of that CoMP set may have sufficient non-CoMP-users. Fig. \ref{fig:iii}(b) illustrates the CoMP-NOMA deployment scenario$-3$ for a two-cell CoMP set, where cell$-2$ does not have any non-CoMP-user to form a NOMA-cluster with the CoMP-users, while cell$-1$ forms a NOMA cluster by grouping both of the CoMP-users and the non-CoMP-user. In this situation, in addition to the requirement of same decoding order for CoMP-users, the decoding order among the CoMP-users itself significantly affects the spectrum efficiency. In the section on numerical results, we will demonstrate the spectral efficiency performance for two different decoding orders of CoMP-users. The achievable throughput formulas for CoMP and non-CoMP users can, however, be expressed similarly as in scenario$-2$.

\section*{Spectral Efficiency Performance of Downlink CoMP-NOMA Systems}
\subsection*{Simulation Assumptions}
In this section, we analyze the  gain in spectral efficiency for downlink CoMP-NOMA systems for different CoMP-NOMA schemes and deployment scenarios discussed earlier  when compared to the CoMP-OMA systems. The  average spectral efficiency (in bits/sec/Hz) is evaluated for all the cells in a CoMP set by using the Shannon's capacity formula. We assume that the BSs do not use sectorization and a BS is located at the center of a circular coverage area. In our proposed CoMP-NOMA systems, to allocate optimal downlink power among the users in a cell, we use the dynamic power allocation model from \cite{msali2016}. It is to be noted that the optimal power allocation  in \cite{msali2016} was derived assuming that  decoding is performed in an ascending oder of channel power gains, that is, a particular user can decode and then cancel all users' signals who have lower channel power gain while all the higher channel gain users act as interferers. In our proposed CoMP-NOMA models, the same CoMP-users would be grouped into multiple NOMA clusters at different CoMP-cells but their decoding order will be same for all clusters. Therefore, the decoding order for a CoMP-user will not follow the the ascending channel gain at all NOMA clusters. For each NOMA cluster, however, we can derive the optimal power allocation solution for the resultant decoding order for CoMP-users by following the same procedure as in \cite{msali2016}. 

In case of orthogonal multiple access (OMA), the transmit power is allocated in proportional to the amount of spectrum resources. The major simulation parameters are as follows: inter-BS distance is $1$ Km, BS transmit power is $43$ dBm, noise spectral density is $-139$ dBm/Hz, system overall bandwidth is $8.64$ MHz,  path-loss coefficient is $4$, the minimum difference between received power (normalized with respect to noise power) of the decoded signal and the non-decoded signal(s) (i.e., $p_{tol}$) is $20$ dBm, and single antenna at BS and UE ends. We also assume that the non-CoMP-users are distributed within the cell-center areas for $400$ m radius in each cell, and the non-CoMP-users of two different cells do not interfere with each other. In addition, in each NOMA cluster, the user who can decode and then cancel all the other users' signals (and hence does not experience any inter-user interference), is referred to as the \textit{cluster-head}.. 

A flat-fading Rayleigh channel having channel power gain with zero mean and unit variance as well as path-loss is considered. For all simulations, the non-CoMP-users are considered at a fixed distance within their distribution areas, while a random distance is considered for CoMP-users outside the non-CoMP-user's coverage areas (measured in terms of the \textit{cell-edge coverage distance}). Perfect channel state information  (CSI) is assumed to be available at the BS ends. All the simulations are done for a single transmission time interval. However, these instantaneous channel gains are averaged over fifty thousands channel realizations. It is also noted that a NOMA cluster achieves maximum throughput gain if the decoding order follows the ascending channel power gain. However, to observe the impact for different decoding orders, we consider that the NOMA cluster in one CoMP-cell follows the ascending channel power  gain decoding, while another CoMP-cell follows a different decoding order while maintaining the same decoding order for CoMP-users. Since the working principles of CB-CoMP-NOMA, and DPS-CoMP-NOMA are the same as that  of the conventional single-cell NOMA, we mainly evaluate the performance for JT-CoMP-NOMA in this section.

\begin{figure}[h]
\begin{center}
	\includegraphics[width=3.4 in]{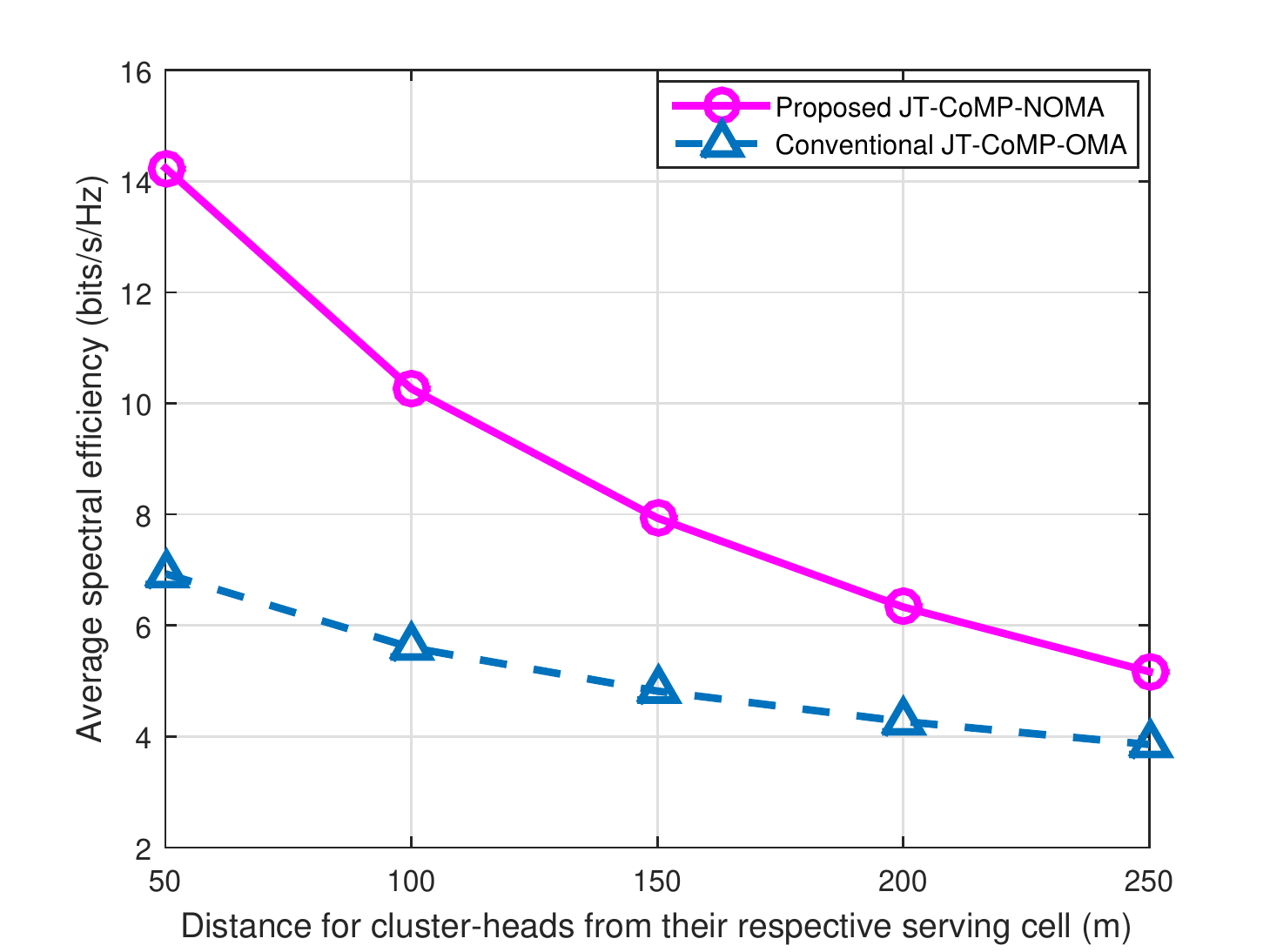}
	\caption{Average spectral efficiency for JT-CoMP-NOMA and JT-CoMP-OMA systems for deployment scenario$-1$ with one CoMP-user and two non-CoMP-users in each CoMP-cell.}
	\label{fig:iv}
 \end{center}
\end{figure} 
\begin{figure}[h]
\begin{center}
	\includegraphics[width=3.4 in]{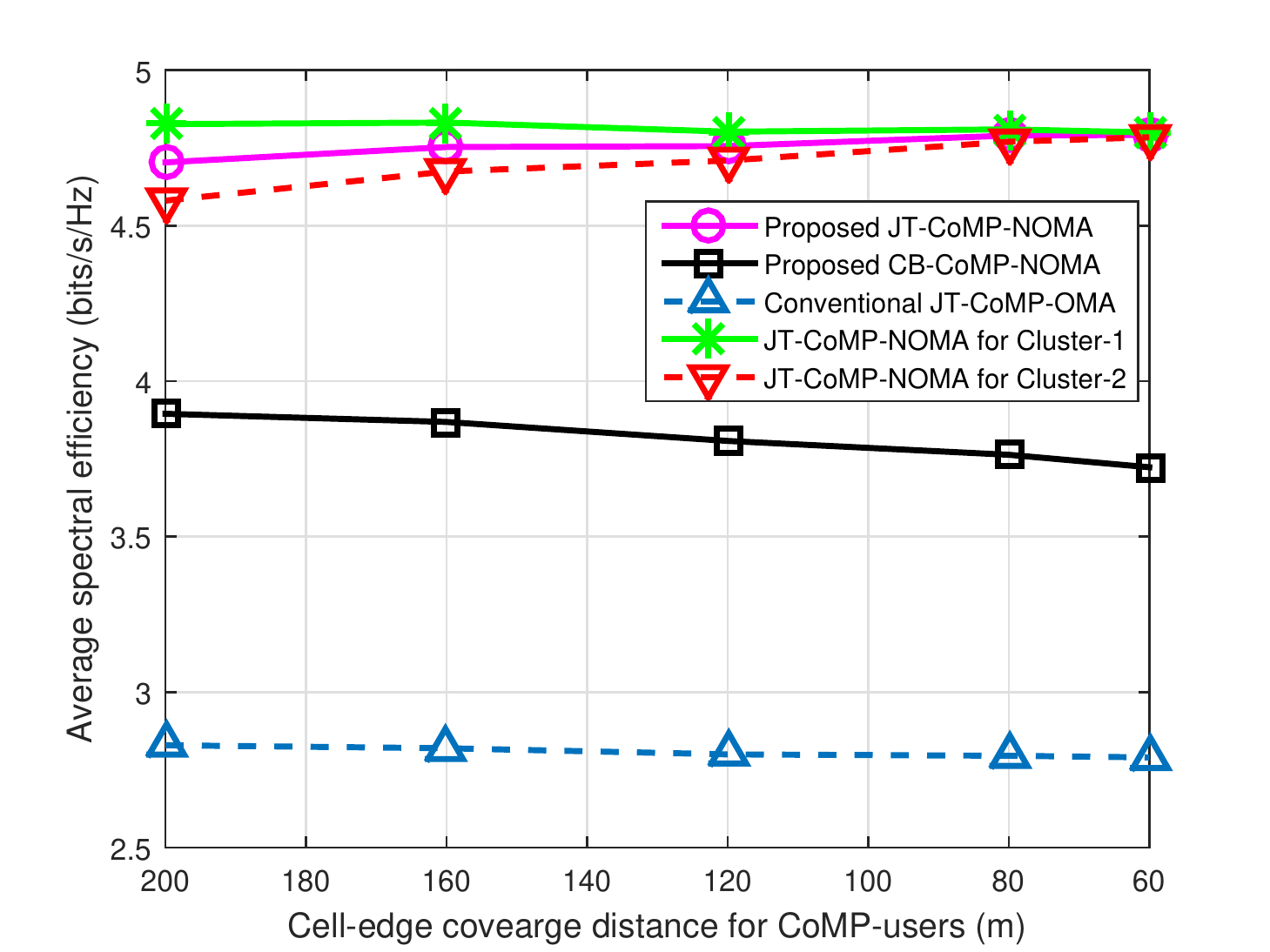}
	\caption{Average spectral efficiency for JT-CoMP-NOMA, CS-CoMP-NOMA and JT-CoMP-OMA systems for deployment scenario$-2$ with two CoMP-users and one non-CoMP-user in each CoMP-cell.}
	\label{fig:v}
 \end{center}
\end{figure} 
\begin{figure}[h]
\begin{center}
	\includegraphics[width=3.4 in]{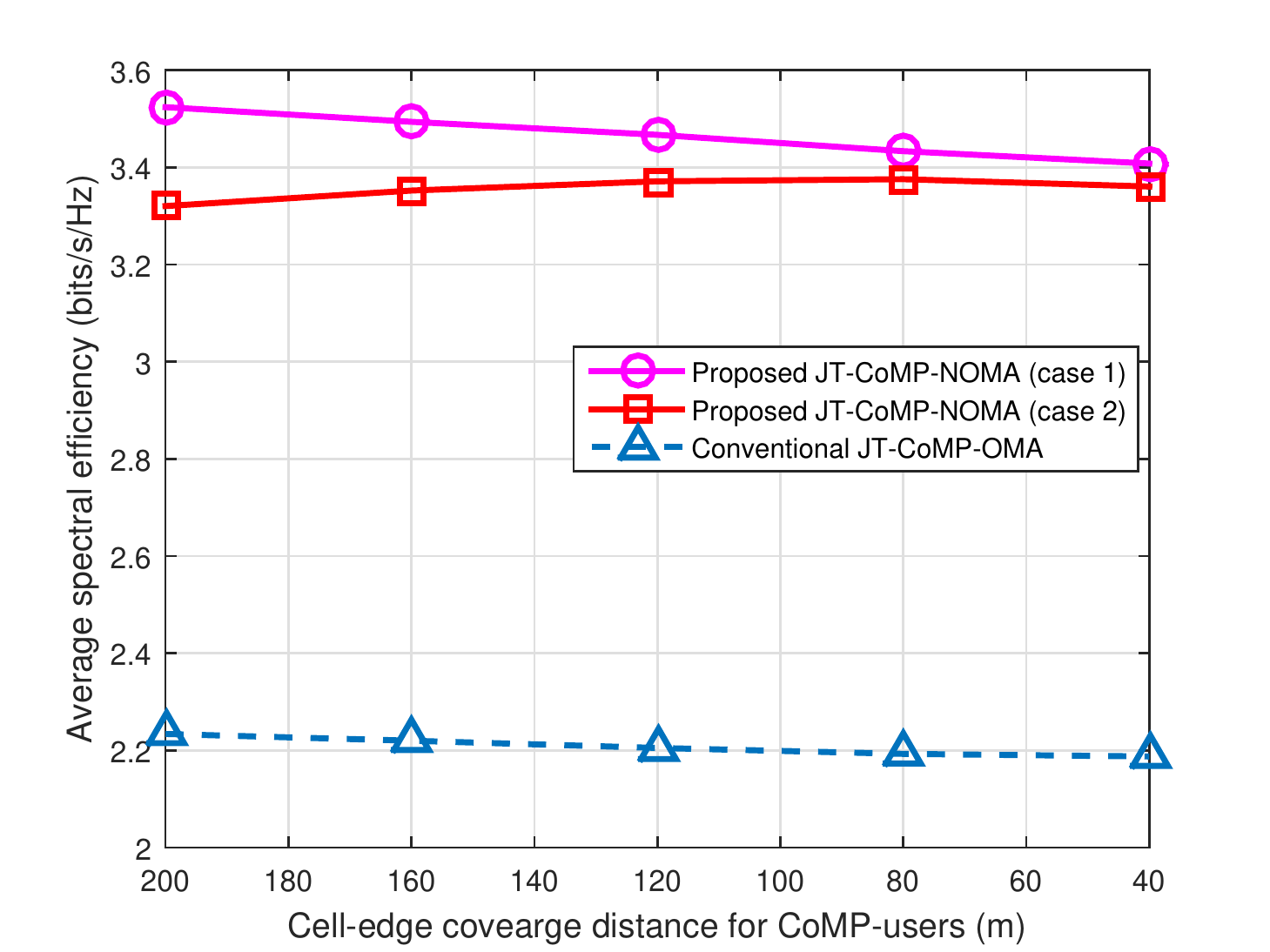}
	\caption{Average spectral efficiency for JT-CoMP-NOMA and JT-CoMP-OMA systems for deployment scenario$-3$ with two CoMP-users and one non-CoMP-user in cell$-1$ but none non-CoMP-user in cell$-2$.}
	\label{fig:vi}
 \end{center}
\end{figure} 

\subsection*{Simulation Results}


We simulate three different models for three aforementioned \textit{deployment scenarios}. Fig. \ref{fig:iv} shows the average spectral efficiency for the proposed JT-CoMP-NOMA and the conventional JT-CoMP-OMA for \textit{deployment scenario$-1$} illustrated in Fig. \ref{fig:iii}(a), where we consider one CoMP-user in a two-cell CoMP set and two non-CoMP-users in each cell. The guaranteed throughput requirement for each user in JT-CoMP-NOMA system is equal to its achievable JT-CoMP-OMA throughput considering equal spectrum allocation. For example, if $B$ be the system bandwidth, for OMA operations, each CoMP-user and non-CoMP-user will be allocated $B/3$ bandwidth from each cell. For the CoMP-user, both the cells allocate the same spectrum resource and transmit the same data stream, thus the receiver performance is improved. The random distance for CoMP-user is considered at $200$ m cell edge radius. The average spectral efficiency is measured for different distance for cluster-heads, while the second non-CoMP-user in each cell is assumed at $300$ m distance from the BS. 

The key observation from Fig. \ref{fig:iv} is that the average spectral efficiency gain of JT-CoMP-NOMA system for \textit{deployment scenario$-1$} is much higher in comparison to that of a JT-CoMP-OMA system, and the performance gain largely depends on the channel gain for cluster-head. Since the cluster-head is the highest channel gain user in each cluster, thus the performance gain is very obvious. In this case, the more distinct channel for cluster-head than the other NOMA users provides more spectral efficiency compare to their OMA counterparts. 

The average spectral efficiency for JT-CoMP-NOMA, CS-CoMP-NOMA, and JT-CoMP-OMA for \textit{deployment scenario$-2$ (illustrated in Fig. \ref{fig:i}(c) for JT-COMP-NOMA}), are shown in Fig. \ref{fig:v}. Here, we consider one non-CoMP-user in each CoMP-cell at a distance $250$ m  from its serving BS, while two CoMP-users are randomly distributed at various cell-edge coverage distance. The guaranteed throughput requirement for each user in JT-CoMP-NOMA and CS-CoMP-NOMA systems is equal to their achievable JT-CoMP-OMA throughput by considering equal spectrum resource allocation. As we mentioned earlier, for JT-CoMP-NOMA, both CoMP-users can use full spectrum resource by forming $3$-user NOMA cluster (two CoMP-users and one non-CoMP-user) at both CoMP-cells, while each CoMP-user can use at most $50\%$ spectrum resources (orthogonal resources) in CB-CoMP-NOMA system by forming $2$-user NOMA cluster (one CoMP-users and one non-CoMP-user). The additional $50\%$ spectrum is allocated only to non-CoMP-user at both cell in CB-CoMP-NOMA system. 

Fig. \ref{fig:v} shows the average spectral efficiency gain for CoMP-NOMA systems over CoMP-OMA systems. It is observed that JT-CoMP NOMA provides a much higher spectral efficiency than CS-CoMP-NOMA. Since there are two CoMP-users in each JT-CoMP-NOMA cluster, the NOMA cluster that uses optimal decoding order (ascending channel gain order) will have a higher spectral efficiency than the other which uses a non-optimal decoding order. In Fig. \ref{fig:v}, we consider that cluster$-1$ formed in cell$-1$ uses the optimal decoding order, while cluster$-2$ formed in cell$-2$ uses a non-optimal decoding order. However, as the cluster-head in both of the JT-CoMP-NOMA clusters is the highest channel gain non-CoMP-user, the variation for spectral efficiency is not significant. 

Fig. \ref{fig:vi} shows the spectral efficiency gain for JT-CoMP-NOMA and JT-CoMP-OMA in \textit{deployment scenario$-3$} illustrated in Fig. \ref{fig:iii}(b), where two CoMP-users are assumed in a two-cell CoMP set, while one non-CoMP-user in cell$-1$ but there is no non-CoMP-user in cell$-2$. Two CoMP-users are randomly distributed at various cell-edge coverage distance, while the non-CoMP-user in cell$-1$ is located at a distance of $250$ m  from the BS. The guaranteed throughput requirement for each user in JT-CoMP-NOMA system is equal to the achievable JT-CoMP-OMA throughput by considering equal spectrum allocation to all users. In the corresponding OMA system, cell$-2$ allocates more spectrum resources ($50\%$) to each CoMP-user than what cell$-1$ allocates ($33.33\%$) but both allocations are in the same band, and thus each  CoMP-user gets  the same message signal over the same $33.33\%$ spectrum resource from both the CoMP-cells. In the additional OMA spectrum resource ($16.67\%$), cell$-2$ sends additional data to the CoMP-users.  

Similar to the two other simulation results, Fig. \ref{fig:vi} also shows significant spectral efficiency gain of JT-CoMP-NOMA system over the OMA system. Since the cluster-head in cell$-2$ is a cell-edge user, the performance gain in Fig. \ref{fig:vi} is not as good as the other two results. As the cluster-head in NOMA cluster$-2$ formed in cell$-2$ is a CoMP-user, the performance gain is better for optimal decoding order for cluster$-2$ (\textit{case 1} in Fig. \ref{fig:vi}). However, this performance gain is not significantly high in comparison to that in the case of optimal decoding order for cluster$-1$ (\textit{case 2} in Fig. \ref{fig:vi}). The reason is that the cluster-head in cluster$-2$ is a cell-edge user (i.e., has a low channel power gain) and the channel gains among the cluster-head and another user in cluster$-2$ is less distinctive, while the cluster-head in NOMA cluster$-1$ is a non-CoMP user with more distinctive channel gain than the other users in that cluster.

\section*{Summary and Open Challenges}

In this article, we have demonstrated the gain in spectral efficiency performance for CoMP transmission in downlink homogeneous multi-cell NOMA systems by considering distributed power allocation. We have identified the necessary conditions required to perform CoMP-NOMA in downlink transmission under distributed power allocation. Different CoMP-NOMA schemes have been numerically analyzed under various network deployment scenarios. All of the simulation results reveal the superior spectral efficiency performance of CoMP-NOMA systems over their counterpart  CoMP-OMA systems. However, among all the CoMP schemes, JT-CoMP-NOMA provides the highest  spectral efficiency gain. This is due to the fact that, all the CoMP-users can use the same spectral resources by forming NOMA clusters at all coordinating cells. On the other hand, orthogonal spectrum resource allocation is required among the CoMP-users in other CoMP-NOMA schemes. 

The requirement for perfect CSI availability at all the coordinating cells is a common challenge  for all CoMP transmission systems while  SIC is the key challenge for a NOMA system to protect the error propagation. In addition, to maximize the overall spectral efficiency in all the coordinating cells in a CoMP-NOMA system and  to implement the CoMP-NOMA in downlink heterogeneous networks (HetNets) and MIMO systems, some additional challenges need to be overcome. The major potential challenges  are as follows:

\begin{itemize}
\item In this article, we have used optimal power allocation for a given decoding order for each NOMA cluster. However, determining the optimal decoding order among all the coordinating cells is a challenging task. An exhaustive search algorithm could be a solution for optimal decoding order but the complexity of such a solution would be very high for a CoMP set with more than two cells and/or two CoMP-users. Finding low-complexity  near-optimal user clustering schemes for CoMP-NOMA systems is an open challenge.

\item When the cluster-head is the highest channel gain user of a NOMA cluster, our optimal power allocation solution for sum-throughput maximization provides minimum power to meet the guaranteed throughput requirement for all NOMA users except the cluster-head who gets all the residual power, by maintaining the SIC decoding requirements. Thus, the sum-throughput would be the maximum for the given minimum throughput requirement of each NOMA user. However, in JT-CoMP-NOMA, each CoMP-user receives the same data stream transmitted over the same spectrum resources from multiple cells, while their channel gains at each coordinating cell are different. Thus how much power to allocate to a JT-CoMP user at each coordinating cell to satisfy the user's rate requirement while achieving the optimal spectral efficiency in all the coordinating cells is another open research challenge. Determining the optimal decoding order and optimal rate requirement for a CoMP-NOMA system is a joint optimization problem.

\item In downlink co-channel HetNets, the small cell users experience strong inter-cell interference from the high power macro-cell. In a NOMA system, it is required for a NOMA user to decode and then cancel (i.e., by using  SIC) signals from  other NOMA users' whose decoding order is prior to this user. Since SIC is performed in the power domain, the co-channel macro-cell interference may make the small cell users unable to perform SIC. Therefore, implementation of NOMA in co-channel downlink HetNets will be very  challenging. However, the use of CoMP could be a potential solution for such NOMA-based HetNets.

\item In HetNets, since multiple small cells underly a macro-cell, a CoMP set may be formed among multiple small cells and one macro cell. In a two-tier HetNet, all users in a small cell could be treated as CoMP-users by the macro cell and the corresponding  small cell, while all users in a small cell cannot be treated as CoMP-users  by another small cell. Therefore, application of CoMP in such a NOMA-based downlink HetNet is a challenging task. The concept of location-aware CoMP  \cite{sakr2014}, in which,  the small cell users close to the small cell BSs are treated as non-CoMP-user, would provide a potential solution to this problem. However, the inter-cell interference for small cell non-CoMP-users would be very high. Therefore, the selection of CoMP-users and non-CoMP-users should be done carefully.

\item In this article, we have only considered single antenna at the BS and UE ends, while the application of multiple-antennas at both ends will need to be investigated. The concept of MIMO-NOMA for a non-CoMP system as introduced in \cite{msali2017}-\cite{higuchi2015} can be utilized in a MIMO-CoMP-NOMA system, but a thorough performance analysis would be required. 
\end{itemize} 



\bibliographystyle{IEEE}

\end{document}